# COMPARISON OF f-Q SCALING IN WINEGLASS AND RADIAL MODES IN RING RESONATORS

*Siddharth Tallur and Sunil A. Bhave*
OxideMEMS Laboratory, Cornell University, Ithaca, NY – USA

## ABSTRACT
Low phase noise MEMS oscillators necessitate resonators with high f-Q. Resonators achieving high f-Q (mechanical frequency-quality factor product) close to the thermo-elastic damping (TED) limit have been demonstrated at expense of feed-through. Here we present a study comparing frequency scaling of quality factors of wineglass and radial modes in a ring resonator using an opto-mechanical two port transmission measurement. Higher harmonics of the wineglass mode show an increasing trend in the f-Q product, as compared to a saturation of f-Q for radial modes. The measured f-Q of $5.11 \times 10^{13}$Hz at 9.82GHz in air at room temperature for a wineglass mode is close to the highest measured values in silicon resonators.

## INTRODUCTION

MEMS resonators designed for oscillator applications to date have quality factors lower than the TED limit [1, 2]. For designing high performance oscillators, two figures of merit that one should maximize are f-Q and $k_t^2$-Q (electromechanical coupling constant-mechanical quality factor product), which dictate the close to carrier phase noise and mechanical energy stored in the resonator respectively. While progress has been made towards optimizing the latter [3, 4], resonators demonstrating high f-Q products close to the TED limit have managed to do so at expense of feed-through, thus necessitating indirect measurement techniques such as LO (local oscillator) mixing [5] or FET-based sense [6].

Opto-mechanical transduction has been previously shown to afford a direct measurement scheme to probe the mechanical modes of MEMS resonators [7]. Indeed, the high sensitivity provided by this scheme makes it possible to detect thermal Brownian motion of these resonators. Here we utilize this measurement scheme in combination with electrostatic air gap actuation to study frequency scaling of quality factors of wineglass modes and radial modes in ring resonators. The next section provides details on the device design and a qualitative comparison of wineglass and radial modes and trends to expect. Then we present experimental results to study these trends and derive insights based on our findings to design higher performance oscillators.

## DEVICE DESIGN
**Device choice**
We choose a ring geometry over a disk as the former combines the ability to achieve high mechanical frequencies and high capacitive actuation area. The modulator utilizes a coupled resonator design to decouple the mechanics and the optics in the opto-mechanical resonator [7]. Wineglass modes of such resonators have been studied previously from theoretical [8] and experimental [5] standpoints. In case of 4-spoke supported released ring resonators, wineglass modes have less stress distribution along the inner perimeter of the ring in the two orthogonal 0°-180° and 90°-270° directions.

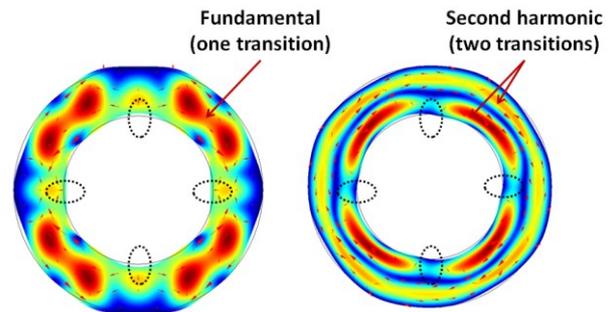

*Figure 1: FEM simulation for mechanical mode shapes of the fundamental and second harmonic of a compound wineglass mode displaying the stress distribution. The encircled points on the inner periphery of the ring resonator correspond to points where the ring will be supported by spokes. The stress at these points is smaller for the second harmonic as compared to the fundamental mode.*

In comparison, radial mechanical modes can never have zero stress distribution at the ring perimeter and thus are expected to have lower mechanical Qs. Moreover higher harmonics of wineglass modes have lesser stress near the spokes and thus should be more immune to anchor loss [9]. An illustration of this is shown in Figure 1, which compares the stress distribution for two compound wineglass modes. The stress in the orthogonal axes at points on the inner periphery of the ring resonator is smaller for the higher harmonic as compared to the fundamental mode. Lower anchor losses at higher frequencies augur well for designing low phase noise oscillators employing these high mechanical Q harmonics.

**Optimizing mechanical and optical performance**
The ring resonator is designed to have a width of 3.8μm and an outer radius of 9.5μm. The coupling spring length is chosen to be 3λ/2 corresponding to the wavelength of the fundamental compound radial mode. This ensures strong coupling for the radial mechanical modes between the two ring resonators. The optimal coupling spring length for the wineglass mode is similar. The ring dimension is chosen to design for a high optical quality factor resonance in the widely used C-band (conventional "erbium window" band, spanning wavelengths from 1,530nm to 1,565nm). FEM simulations using perfectly matched layers (PML) [10] are used to determine the optical quality factor of the optical whispering gallery mode (WGM). Figure 2 below shows a transverse electric (TE) mode for the opto-mechanical resonator at 1,558.03nm with an intrinsic optical Q of 75,150.

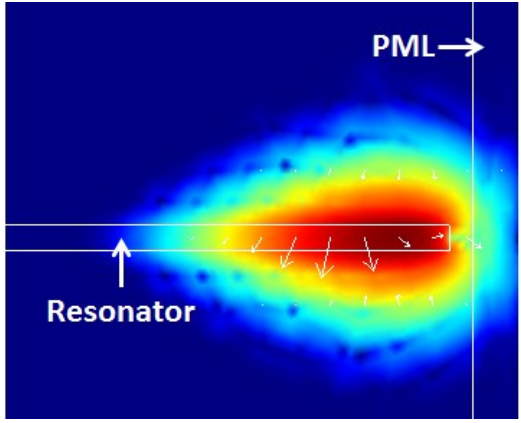

*Figure 2: FEM simulation for optical mode profile of the silicon opto-mechanical resonator. The axisymmetric PDE mode simulation in COMSOL 3.5 [10] is set up to solve Maxwell's equations for a cross section of the ring resonator to obtain the profile of the optical mode. The simulated optical resonance wavelength is 1,558.03nm and intrinsic optical Q is 75,150.*

The modulator is designed in a single-crystal-silicon (SCS) on insulator (SOI) platform following the fabrication process developed in [7]. Figure 3 shows a scanning electron micrograph (SEM) of the device. The device thickness is 220nm, the ring-waveguide gap is 80nm, and the ring-electrode gap is 130nm.

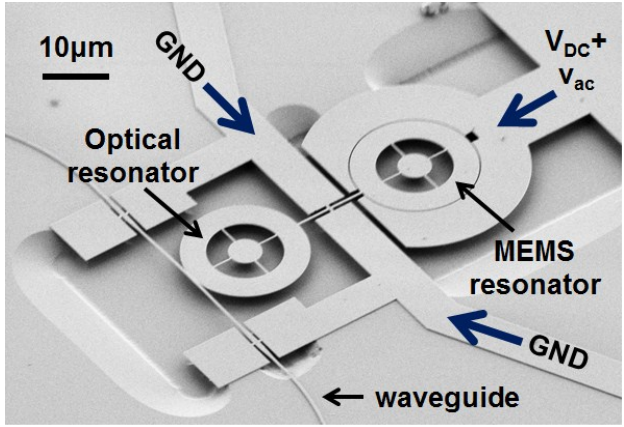

*Figure 3: Scanning electron micrograph of the coupled silicon opto-mechanical resonator. The ring resonators have a width of 3.8μm and outer radius 9.5μm. The ring-waveguide gap is 80nm, and the ring-electrode gap is 130nm.*

## EXPERIMENTAL RESULTS
### Optical characterization

We couple light from a tunable laser into the grating coupler and monitor the output using an optical power meter. The laser wavelength is swept to identify a high Q optical resonance. Figure 4 shows an optical resonance with a total loaded optical quality factor of 31,300 and an intrinsic optical Q ~ 90,000. The grating couplers introduce a loss of 8dB per coupler.

### Characterization of mechanical modes

We use the setup illustrated in Figure 5 to study the mechanical modes of the modulator.

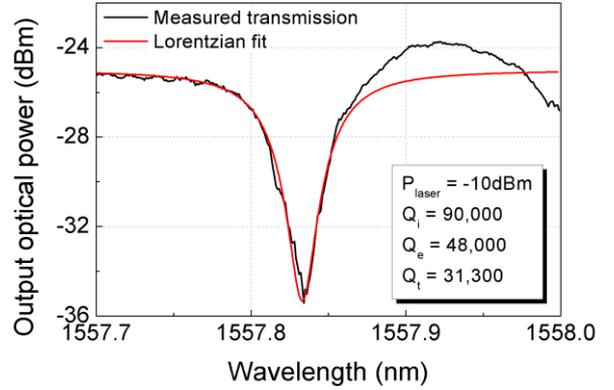

*Figure 4: Optical spectrum showing the high-Q resonance. The input laser power was -10dBm.*

We use Newport 1544-A as the photo-detector in our system. This detector has a bandwidth of 15GHz. Built-in InGaAs PIN photodiode and a GaAs HBT amplifier offer high responsivity at 1,557nm thus enabling a clean frequency response measurement. An Agilent Agilent N5230A PNA-L Network Analyzer in a 2-port configuration is used to study the transmission spectrum for the modulator.

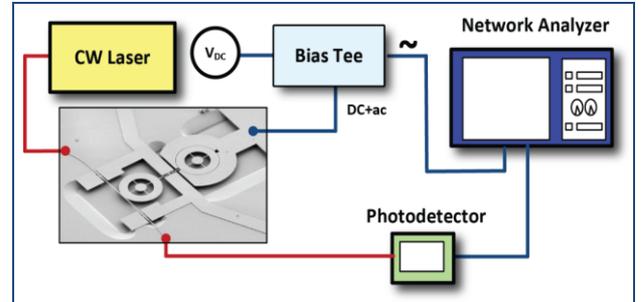

*Figure 5: Experimental setup for characterization of the RF transmission of the resonator. Mechanical modes are actuated via capacitive air-gap drive (DC bias and AC voltage from port 1 of network analyzer) and sensed via optical intensity modulation detected using a Newport 1544-A photo receiver, on port 2 of the network analyzer.*

Figure 6 shows the measured transmission spectrum for the modulator measured in air at room temperature. Mechanical motion is driven by applying a combination of 20V DC bias and 0dBm RF power at the electrodes. Various mechanical modes of the modulator are detected. For the purpose of this study, we focus on harmonics of the compound radial mode (highlighted with red circles) and the compound wineglass mode (highlighted with green triangles). Due to the small size, e-beam defined geometry and direct displacement, we can match resonance peaks in the transmission spectrum to COMSOL modal simulations, all the way to 10GHz. The output signal power at higher frequencies is limited by the detector sideband (12GHz), the transduction efficiency (capacitive air gap actuation), and the intensity modulation efficiency, which is governed by the optical Q factor [11].

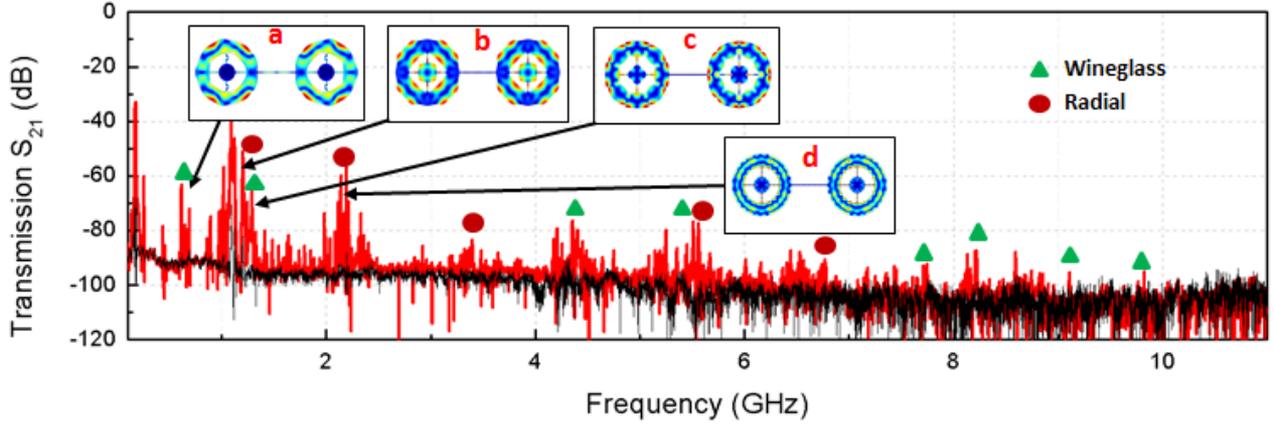

*Figure 6: Wide RF transmission spectrum of the resonator. Various mechanical modes of the device are sensed via optical intensity modulation at the photo receiver. A combination of 0dBm RF power and 20V DC bias was applied at the electrodes. Insets: COMSOL simulations of mode-shapes for fundamental and second harmonic of the wineglass mode (a,c) and radial mode (b,d). The black curve corresponds to 0V DC bias.*

We measure the quality factor for the different harmonics of the radial and wineglass modes. Figure 8 shows the variation of f-Q for the radial mode and the wineglass mode families. The Q for radial mode decreases for higher harmonics in such a way that the measured f-Q saturates to a value close to $9 \times 10^{12}$Hz. In contrast to the radial mode, the f-Q for the wineglass mode family increases for higher harmonics. This can be attributed to reduced anchor losses on account of smaller displacements at the spokes. The highest f-Q is measured to be $5.11 \times 10^{13}$Hz at 9.82GHz.

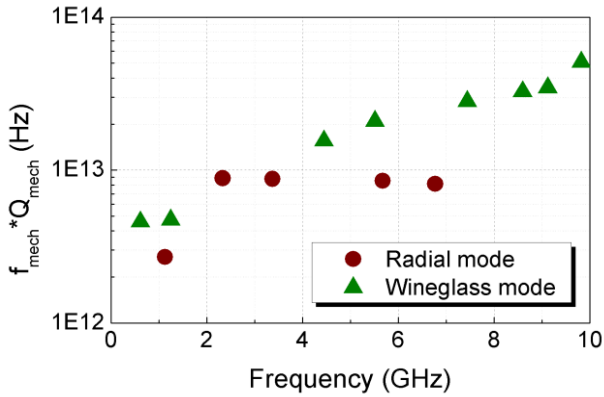

*Figure 7: f-Q vs f plotted for harmonics of the radial mode and the wineglass mode. Unlike the radial mode, the f-Q for the wineglass mode does not saturate and the value for higher harmonics keeps increasing. Measurements were carried out at room temperature and atmospheric pressure.*

**Temperature dependence of quality factors**

Based on the measurements we perform at room temperature, we believe the quality factors for the radial modes are limited by anchor losses. To corroborate this hypothesis, we investigated the variation of quality factors for the wineglass and radial mode families with temperature. The device was tested in a Lakeshore TTP4 probe station under 9μTorr pressure, and liquid nitrogen was used to cool the chamber. A modified optical probe arm was introduced in the probe station to couple laser light to the devices via grating couplers.

We measured the performance of this modulator at various temperatures down from room temperature to 80K. The TTP4 optical probe arm connectors introduce further losses in addition to the losses due to grating couplers. This limits the highest observable signal frequency to roughly 6GHz, and signals at frequencies beyond this value are buried under noise.

Figure 8 shows the variation of f-Q as a function of temperature for harmonics of the radial mode. The saturation in the value of f-Q for higher harmonics seen at room temperature no longer holds and the variation shows a rising trend. It would be misleading to interpret this as transition from anchor-loss dominated regime to phonon-phonon dissipation limit. The measured value of f-Q at 5.67GHz at 80K is 10X lower than the maximum calculated limits for <100> silicon [13].

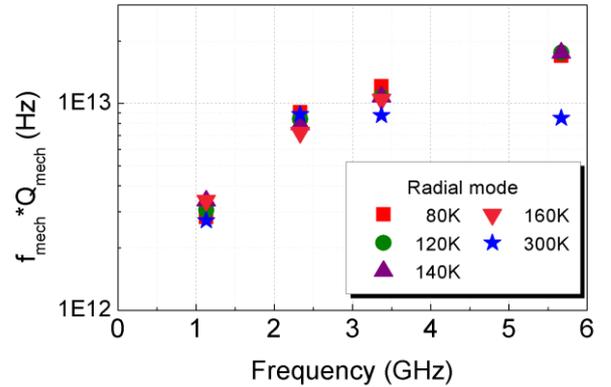

*Figure 8: f-Q vs f plotted for harmonics of the radial mode at various temperatures. At room temperature the f-Q for higher harmonics of the radial mode saturates at a value close to $9 \times 10^{12}$Hz. At lower temperatures the value of f-Q shows an increasing trend.*

The wineglass mode family on the other hand, shows an increasing trend in the f-Q product for higher harmonics at all temperatures. As in the case of radial modes, the values measured are far lesser than the theoretical limit [13]. This implies that higher harmonics of the wineglass mode are more resilient to anchor losses.

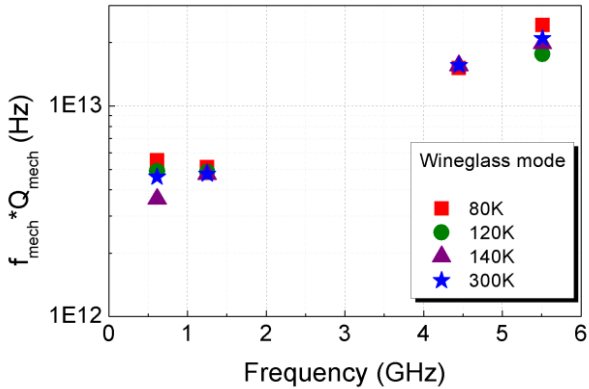

*Figure 9: f-Q vs f plotted for harmonics of the wineglass mode at various temperatures. The f-Q for higher harmonics shows a rising trend at all frequencies.*

## CONCLUSION

In summary, we have characterized and compared the f-Q for radial and wineglass modes of a ring resonator using an integrated electrostatic drive and opto-mechanical sense scheme. The measured f-Q for higher harmonics of the wineglass mode increases with the mode order, indicating anchor loss mitigation. The measured f-Q of $5.11 \times 10^{13}$ Hz at 9.8GHz is comparable to earlier measured highest f-Q values in silicon [6, 12] and diamond [5]. Measurements carried out at lower temperature are indicative that the quality factors of our resonator are limited by anchor losses. A key insight from this study for designing a higher performance oscillator, is to design for wineglass modes and scale to higher frequencies, as long as the transducer remains efficient.

## ACKNOWLEDGEMENTS

This work was supported by the DARPA/MTO's ORCHID program and Intel Academic Research Office. The devices were fabricated at the Cornell NanoScale Science and Technology Facility.

## CONTACT

*S. Tallur; sgt28@cornell.edu